\documentclass[aps]{revtex4}
\usepackage{amsmath}  
\usepackage[dvips]{graphics}

\begin{document}
\title
{
Dynamical Structure Factors of the $S=1/2$ Bond-Alternating Spin Chain with a Next-Nearest-Neighbor Interaction in Magnetic Fields
}

\author
{ 
Mamoru Usami\footnote{Present address: Research and Development Department, Tochigi Nikon Corporation, 770 Midori, Ohtawara 324-8625.} and Sei-ichiro Suga
}

\affiliation
{
Deparatment of Applied Physics, Osaka University, Suita, Osaka 565-0871, Japan
}

%\date
%{
%\today
%}

\begin{abstract}
The dynamical structure factor of the $S=1/2$ bond-alternating spin chain with a next-nearest-neighbor interaction in magnetic field is investigated using the continued fraction method based on the Lanczos algorithm. 
When the plateau exists on the magnetization curve, the longitudinal dynamical structure factor shows a large intensity with a periodic dispersion relation, while the transverse one shows a large intensity with an almost dispersionless mode. The periodicity and the amplitude of the dispersion relation in the longitudinal dynamical structure factor are sensitive to the coupling constants. 
The dynamical structure factor of the $S=1/2$ two-leg ladder in magnetic field is also calculated in the strong interchain-coupling regime. 
 The dynamical structure factor shows gapless or gapful behavior depending on the wave vector along the rung.  
\end{abstract}

\maketitle

%%%%%%%%%%%%%%%%%%%%%%%%%%%%%%%%%%%%%%%%%%%%%%%%%%%%%
\section{Introduction}
%%%%%%%%%%%%%%%%%%%%%%%%%%%%%%%%%%%%%%%%%%%%%%%%%%%%%%
One-dimensional quantum spin systems have attracted great attention both theoretically and experimentally for a long time. 
Recently, Tonegawa {\it et al.} discovered numerically that the $S=1/2$ bond-alternating spin chain with a next-nearest-neighbor (NNN) interaction has a parameter region where the plateau exists on the magnetization curve at half of the saturation value.~\cite{TNK98} 
The wave function of the lowest-energy state in the plateau region was shown to have the periodicity $n=4$,~\cite{TNK98} indicating that spontaneous translational-symmetry breaking, because the periodicity of the Hamiltonian is $n=2$. This result is consistent with the necessary condition for the appearance of the magnetization plateau at half of the saturation value in the $S=1/2$ spin chain~\cite{OYA97},  which indicates that the periodicity of the lowest-energy state must be a multiple of $4$. 
On the basis of bosonization technique combined with the renormalization-group theory, the magnetization process of this system was investigated in a viewpoint of the stability of the magnetization plateau.~\cite{Totsuka98} 
A simple explanation for the two-fold degenerate ground-state in the plateau region was also presented.~\cite{Totsuka98}

The dynamical structure factor (DSF) of quantum spin systems provides us with important information about the intensity of the magnetic response as a function of energy and momentum transfer. 
The DSF of $S=1/2$ one-dimensional quantum spin systems in magnetic fields was measured by inelastic neutron-scattering experiments.~\cite{Endoh,Coldea} 
The experimental results were compared with numerical results obtained by the exact diagonalization method for the finite system,~\cite{IS,Mull,Wup1} and it has been argued that both results are in good agreement. 

The DSF of $S=1/2$ one-dimensional spin-gapped systems in magnetic fields was investigated by numerical calculation~\cite{Wup2,Poil,MH}. Using the recursion method, the longitudinal DSF of the $S=1/2$ Heisenberg chain with a NNN interaction was calculated at the half of the saturated magnetization, and the characteristic distribution of the intensity has been discussed~\cite{Wup2}. The longitudinal DSF of the $S=1/2$ bond-alternating spin chain with a NNN interaction was calculated using the numerical diagonalization method~\cite{Poil,MH}. When some adequate parameters are used, this model is considered to be an effective model Hamiltonian for ${\rm CuGeO_3}$.~\cite{RD,CCE} It has been shown that the large intensity lies at the soft mode,~\cite{Poil,MH} and the distribution of the intensity is remarkably affected by the spin-phonon coupling.~\cite{MH} 
Using bosonization technique, the DSF of the $S=1/2$ one-dimensional spin-gapped systems such as the Heisenberg chain with a NNN interaction, the bond-alternating chain and the two-leg ladder~\cite{FZ99} was calculated systematically in magnetic fields~\cite{CG}. On the basis of the results, Chitra and Giamarchi have concluded that the features of $S=1/2$ one-dimensional spin-gapped systems appear not in thermodynamic quantities but in critical and dynamical properties of the spin correlation function, when the energy gap is collapsed by magnetic fields.~\cite{CG} 

In this paper, we calculate the DSF of the $S=1/2$ bond-alternating spin chain with a NNN interaction in magnetic fields, using the continued fraction method based on the Lanczos algorithm.~\cite{GB87} This model can be regarded as the $S=1/2$ two-leg spin ladder with a diagonal interaction. 
In \S \ref{sec:MODEL}, we summarize the numerical method to make this paper self-contained. In \S \ref{sec:DSFofCHAIN}, we show the DSF of the $S=1/2$ bond-alternating spin chain with a NNN interaction, turning our attention on the behavior at the magnetization plateau. In \S \ref{sec:DSFofLADDER}, the results for the DSF of the $S=1/2$ two-leg spin ladder and the $S=1/2$ two-leg spin ladder with a diagonal interaction are shown at half of the saturated magnetization. We summarize the paper in \S \ref{sec:Summary}.

%%%%%%%%%%%%%%%%%%%%%%%%%%%%%%%%%%%%%%%%%%%%%%%%%%%%%%
\section{Model and numerical method}
\label{sec:MODEL}
%%%%%%%%%%%%%%%%%%%%%%%%%%%%%%%%%%%%%%%%%%%%%%%%%%%%%%
Let us consider the following model Hamiltonian for the $S=1/2$ bond-alternating spin chain with a NNN interaction in magnetic fields, 
%
%%%%%%%%%%%%%%%%%%%%%%%%%%%%%%%%%%%%%%%%%%%%%%%%%%%%%%%%%%%
\begin{eqnarray}
{\cal H} &=& {\cal H}_0 + {\cal H}_z, \label{eq:hamil} \\
{\cal H}_0 &=& 
  J_1 \sum_j
  \mbox{\boldmath $S$}_{1,j} \cdot \mbox{\boldmath $S$}_{2,j}
+ J_2 \sum_j
  \mbox{\boldmath $S$}_{2,j} \cdot \mbox{\boldmath $S$}_{1,j+1},
+ J_3 \sum_{l,j}
  \mbox{\boldmath $S$}_{l,j} \cdot \mbox{\boldmath $S$}_{l,j+1},
\label{eq:hamil_0} \\
{\cal H}_z &=& -MH, \label{eq:hamil_z}
\end{eqnarray}
%%%%%%%%%%%%%%%%%%%%%%%%%%%%%%%%%%%%%%%%%%%%%%%%%%%%%%%%%%%
%
where $\mbox{\boldmath $S$}_{l,j}$ is the $S=1/2$ spin operator on the site $j$ of the sublattice $l$ ($l=1$ or $2$), $J_1$ and $J_2$ are the alternating nearest-neighbor interactions, and $J_3$ is the next-nearest-neighbor interaction. The chain is divided into two sublattices, which consist of every other sites.  The distance between neighboring spins is set equal to unity. 
The magnetization is given by $M=\sum_{l,j} S_{l,j}^z$. We set $J_1, J_2$ and $J_3>0$. The periodic boundary condition is applied. 
Note that the system described by the model Hamiltonian (\ref{eq:hamil}) can be regarded as the $S=1/2$ two-leg spin ladder with a diagonal interaction, where $J_1$, $J_2$ and $J_3$ are the interchain, diagonal and intrachain interactions, respectively.

To calculate the DSF, we first consider the ladder geometry. In this case, the DSF is defined as, 
%
%%%%%%%%%%%%%%%%%%%%%%%%%%%%%%%%%%%%%%%%%%%%%%%%%%%%%%%%%%%
\begin{eqnarray}
S^\mu (q_x,q_y,\omega)
&=& \sum_n | \langle \Psi_n | S_{q_x,q_y}^\mu |
\Psi_0 \rangle |^2 \delta ( \omega - (E_n - E_0) ) \label{eq:IM}\\
&=& - \frac{1}{\pi} \lim_{\varepsilon \rightarrow +0}
\mbox{Im} \langle \Psi_0 | {S_{q_x,q_y}^\mu}^\dagger
\frac{1}{z + E_0 - {\cal H}}
S_{q_x,q_y}^\mu | \Psi_0 \rangle , \label{eq:CFM}
\end{eqnarray}
%%%%%%%%%%%%%%%%%%%%%%%%%%%%%%%%%%%%%%%%%%%%%%%%%%%%%%%%%%%
%
where $z=\omega + {\rm i} \varepsilon$, $\Psi_n$ is the eigenfunction of ${\cal H}$ whose eigenvalue is $E_n$, $\Psi_0$ is the ground state eigenfunction with the eigenvalue $E_0$, and $q_x$ and $q_y$ are the wave vectors along the chain and the rung, respectively. 
The Fourier transform of the spin operator $S _{q_x,q_y}^\mu$ is given by 
%
%%%%%%%%%%%%%%%%%%%%%%%%%%%%%%%%%%%%%%%%%%%%%%%%%%%%%%%%%%%
\begin{equation}
S_{q_x,q_y}^\mu = \frac{1}{\sqrt{N}} \sum_{l,j} 
\exp[{\rm i}(q_x j + q_y l)] S_{l,j}^\mu,
\label{eq:Fourier}
\end{equation}
%%%%%%%%%%%%%%%%%%%%%%%%%%%%%%%%%%%%%%%%%%%%%%%%%%%%%%%%%%%
%
where $\mu=x,y,z$ or  $+,-,z$. 

We calculate $\Psi_0$ numerically in the subspace with a given magnetization. 
The DSF can be expressed by the continued fraction in the tridiagonal basis of the Hamiltonian with the initial vector 
$| \phi_0 \rangle = S_{q_x,q_y}^\mu | \Psi_0 \rangle / 
\sqrt{\langle \Psi_0 | {S_{q_x,q_y}^\mu}^\dagger S_{q_x,q_y}^\mu
| \Psi_0 \rangle }$;~\cite{GB87} 
%
%%%%%%%%%%%%%%%%%%%%%%%%%%%%%%%%%%%%%%%%%%%%%%%%%%%%%%%%%%%
\begin{eqnarray}
S^{\mu} (q_x,q_y,\omega)=-\frac{1}{\pi} \lim_{\varepsilon \rightarrow +0}
\dfrac{\langle \Psi_0 | {S_{q_x,q_y}^\mu}^\dagger S_{q_x,q_y}^\mu | \Psi_0 \rangle}{z-\alpha_0-
	\dfrac{{\beta_1}^2}{z-\alpha_1-
	\dfrac{{\beta_2}^2}{z-\alpha_2-
	\dfrac{{\beta_3}^2}{z-\alpha_3-
	\dfrac{{\beta_4}^2}{z-\alpha_4- \cdots
}}}}}, 
\end{eqnarray}
%%%%%%%%%%%%%%%%%%%%%%%%%%%%%%%%%%%%%%%%%%%%%%%%%%%%%%%%%%%
%
where $\alpha_i$ and $\beta_{i+1}$ $(i=0,1,2, \cdots)$ are the diagonal $(i+1,i+1)$ and the sub-diagonal $(i+2,i+1)$ elements of the tridiagonal Hamiltonian. The matrix elements $\alpha_i$ and $\beta_i$ are calculated by the Lanczos method. We evaluate the continued fraction by the modified Lenz method~\cite{NR}. 
This method enables us to estimate the convergency quantitatively. 
A small but finite $\varepsilon$ is used instead of taking the limit 
$\varepsilon \rightarrow 0$. We set $\varepsilon = 3.0 \times 10^{-2}$. 

In magnetic field along the $z$ axis, rotational symmetry around the $x$ and $y$ axes is broken, while rotational symmetry around the $z$ axis remains. 
Hence, $S^{x} (q_x, q_y, \omega) = S^{y} (q_x, q_y, \omega) \neq S^{z} (q_x, q_y, \omega)$. 
We calculate $S^{+} (q_x, q_y, \omega)$ and $S^{-} (q_x, q_y, \omega)$ to obtain $S^{x} (q_x, q_y, \omega)$ by using $S^{x} (q_x, q_y, \omega) = (1/4) [S^{+} (q_x, q_y, \omega) + S^{-} (q_x, q_y, \omega)]$. 
We now define the magnetic field where the magnetization changes from $M$ to $M+1$ or from $M-1$ to $M$ as 
%
%%%%%%%%%%%%%%%%%%%%%%%%%%%%%%%%%%%%%%%%%%%%%%%%%%%%%%%%%%%
\begin{eqnarray}
E_L(N,M+1)-E_L(N,M) &=& H_1, \label{eq:jiba1} \\
E_L(N,M)-E_L(N,M-1) &=& H_2  \label{eq:jiba2},
\end{eqnarray}
%%%%%%%%%%%%%%%%%%%%%%%%%%%%%%%%%%%%%%%%%%%%%%%%%%%%%%%%%%%
%
where $E_L(N,M)$ is the lowest energy of ${\cal H}_0$ with $N$ spins and the magnetization $M$. 
When we calculate the DSF on the magnetization plateau, we have to fix the magnetic field in the plateau region. We use $H_1$ for $S^{-} (q_x, q_y, \omega)$ and $H_2$ for  $S^{+} (q_x, q_y, \omega)$, indicating that $S^{-} (q_x, q_y, \omega)$ and $S^{+} (q_x, q_y, \omega)$ are calculated at the highest and lowest ends of the magnetic field on the magnetization plateau, respectively. Then, we fix the magnetic field $H_P$ in $H_2<H_P<H_1$. 
Shifting $S^{-} (q_x, q_y, \omega)$ and $S^{+} (q_x, q_y, \omega)$ along the $\omega$ axis by $H_1-H_P$ and $H_P-H_2$, respectively, and adding up them thus obtained, we can calculate $S^{x} (q_x, q_y, \omega) \left( = (1/4)[S^{+} (q_x, q_y, \omega) + S^{-} (q_x, q_y, \omega)] \right)$ in a given magnetic field $H_P$. 
When the magnetization plateau does not appear, the finite size scaling based on conformal field theory is worthwhile to calculate the precise value of the magnetic field in the thermodynamic limit for a given magnetization $m=N/M$.~\cite{ST91} Nevertheless, we take the average of $H_1$ and $H_2$ to determine the magnetic field for a given $m$, because the error in the magnetic field thus obtained shifts slightly the DSF along the $\omega$ axis.

For the one-dimensional geometry, we do not have to distinguish $q_y$ from $q_x$. Therefore, (\ref{eq:IM}) and (\ref{eq:Fourier}) are simply described as 
%
%%%%%%%%%%%%%%%%%%%%%%%%%%%%%%%%%%%%%%%%%%%%%%%%%%%%%%%%%%%
\begin{eqnarray}
S^\mu (q, \omega)
&=& \sum_n | \langle \Psi_n | S_q^\mu |
\Psi_0 \rangle |^2 \delta ( \omega - (E_n - E_0) ), \label{eq:IM1} \\
S_q^{\mu} &=& \frac{1}{\sqrt{N}} \sum_{l,j}
\exp [{\rm i}(2j+l)q] S_{l,j}^{\mu}. \label{eq:Fourier1}
\end{eqnarray}
%%%%%%%%%%%%%%%%%%%%%%%%%%%%%%%%%%%%%%%%%%%%%%%%%%%%%%%%%%%
%
On the basis of (\ref{eq:IM1}) and (\ref{eq:Fourier1}), the calculation of the DSF for the one-dimensional geometry is performed in the same way as done for the ladder geometry.

%%%%%%%%%%%%%%%%%%%%%%%%%%%%%%%%%%%%%%%%%%%%%%%%%%%%%%
\section{DSF of the $S=1/2$ bond-alternating spin 
chain with a NNN interaction}
\label{sec:DSFofCHAIN}
%%%%%%%%%%%%%%%%%%%%%%%%%%%%%%%%%%%%%%%%%%%%%%%%%%%%%%
%
%
%%%%%%%%%%%%%%%%%%%%%%%%%%%%%%%%%%%%%%%%%%%%%%%%%%%%%%%%%%%
\begin{figure}[h]
\begin{center}
\includegraphics{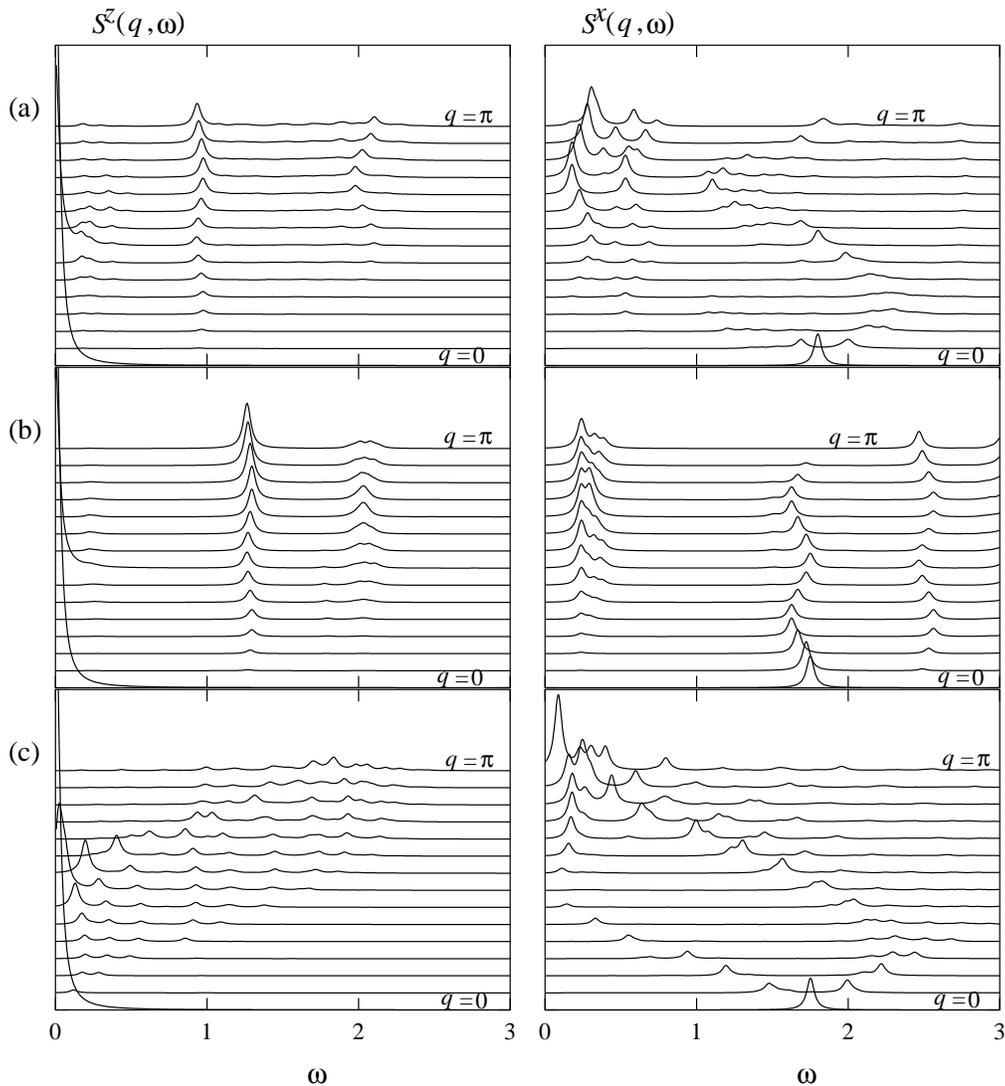}
%\epsfile{file=figure1,height=15cm}
%\figureheight{15cm}
\end{center}
\caption{The DSF of the $S=1/2$ bond-alternating spin chains with a NNN interaction in magnetic fields is shown for $m=1/4$. 
Parameters are 
(a) $J_1=1.2, J_2=0.8, J_3=0.4$, 
(b) $J_1=1.5, J_2=0.5, J_3=0.3$, and 
(c) $J_1=1.1, J_2=0.9, J_3=0.3$. 
The magnetic fields for $m=1/4$ are 
(a) $H=1.80$, (b) $H=1.75$, and (c) $H \sim 1.751$.
In (a) and (b), the plateau exists on the magnetization curve.}
\label{fig:chain}
\end{figure}
%%%%%%%%%%%%%%%%%%%%%%%%%%%%%%%%%%%%%%%%%%%%%%%%%%%%%%%%%%%
%
The DSF of the $S=1/2$ bond-alternating spin chains with a NNN interaction is shown in Fig.~\ref{fig:chain} for $m=1/4$ with $(N,M)=(28,7)$. 
Note that all figures in this paper have the same scale concerning the intensity of the DSF. 
Parameters are 
(a) $J_1=1.2, J_2=0.8, J_3=0.4$, 
(b) $J_1=1.5, J_2=0.5, J_3=0.3$, and 
(c) $J_1=1.1, J_2=0.9, J_3=0.3$. 
According to ref. 1, the plateau exists on the magnetization curve in Figs.~\ref{fig:chain}(a) and (b), but do not exist in Fig.~\ref{fig:chain}(c). 
The magnetic fields for $m=1/4$ are 
(a) $H=1.80$, (b) $H=1.75$, and (c) $H \sim 1.751$, respectively. 
In Figs.~\ref{fig:chain}(a) and (b), two divergency of $S^z (q, \omega)$ in $\omega \rightarrow 0$ are seen at $q=0$ and $q= \pi/2$, whereas in  Fig.~\ref{fig:chain}(c) a divergence of $S^z (q, \omega)$ in $\omega \rightarrow 0$ is seen only at $q=0$. The results for $S^z (q, \omega)$ indicate that in the systems of Figs.~\ref{fig:chain}(a) and (b), translational symmetry is broken and the periodicity is twice as large as that in Fig.~\ref{fig:chain}(c). Thus, the condition for the appearance of the magnetization plateau~\cite{OYA97} is satisfied. 
The peaks of $S^z (q, \omega)$ around $\omega \sim 1.0$ in Fig.~\ref{fig:chain}(a) and $\omega \sim 1.3$ in Fig.~\ref{fig:chain}(b) are almost dispersionless. 
On the contrary, the results for $S^x(q,\omega)$ change qualitatively depending on the parameters as shown in Figs.~\ref{fig:chain}(a) and (b). 
In Fig.~\ref{fig:chain}(a), there are small peaks around $\omega \sim 1.7$. The dispersion relation exhibits cosine-like $q$ dependence with the periodicity $\pi$ and the amplitude $\sim$ $0.7$.  
In Fig.~\ref{fig:chain}(b), the peaks of $S^x (q, \omega)$ exhibit sine-like $q$ dependence with the periodicity $\pi/2$ and the small amplitude $\sim$ $0.1$. 

%
%%%%%%%%%%%%%%%%%%%%%%%%%%%%%%%%%%%%%%%%%%%%%%%%%%%%%%%%%%%
\begin{figure}[h]
\begin{center}
\includegraphics{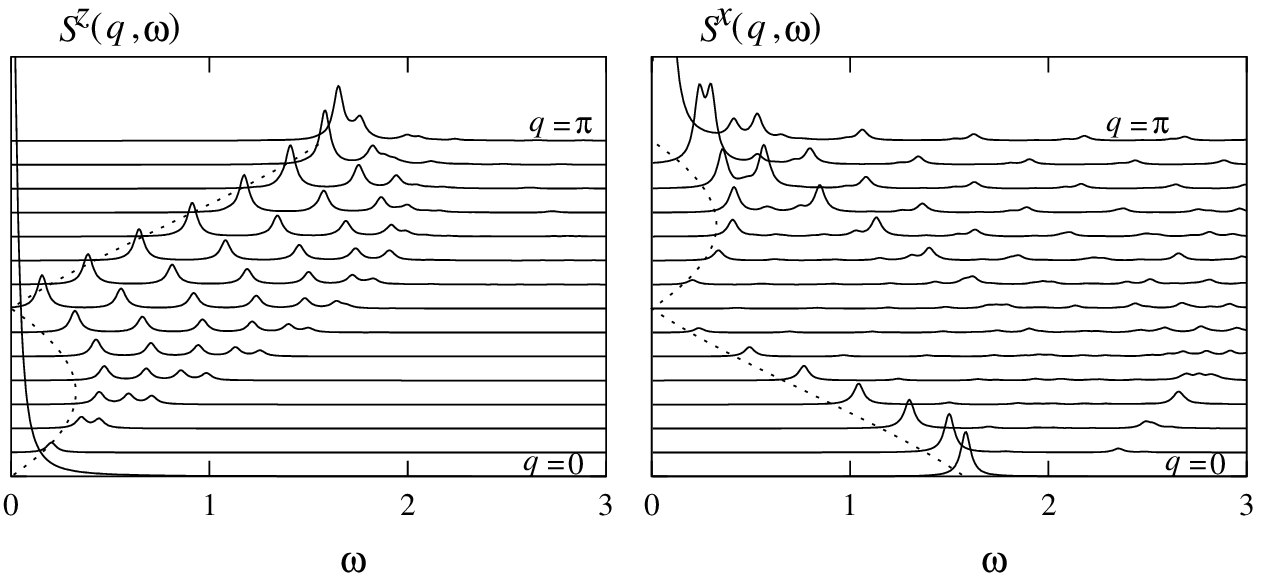}
%\epsfile{file=figure2,height=7cm}
%\figureheight{7cm}
\end{center}
\caption{The DSF of the $S=1/2$ Heisenberg chain in magnetic fields is shown for $m=1/4$. Parameters are $J_1=J_2=1.0$ and $J_3=0$.
}
\label{fig:heisenberg}
\end{figure}
%%%%%%%%%%%%%%%%%%%%%%%%%%%%%%%%%%%%%%%%%%%%%%%%%%%%%%%%%%%
%

We compare the DSF in the absence of the magnetization plateau shown in Fig.~\ref{fig:chain}(c) with that of the $S=1/2$ Heisenberg chain ($J_1=J_2=1.0, J_3=0$) for $m=1/4$ ($(N,M)=(28,7)$) shown in Fig.~\ref{fig:heisenberg}. 
The magnetic field for $m=1/4$ in the latter system is evaluated as $H \sim 1.584$, which is close to the exact value $\sim1.597$~\cite{Mull,Griffiths64}.
The dotted lines denote the lower bound of the excitation continuum obtained from the Bethe ansatz solution~\cite{IS77}.  
The results for $S^z (q, \omega)$ and $S^x (q, \omega)$ in both systems are qualitatively the same, although the additional NNN interaction exists in the system shown in Fig.~\ref{fig:chain}(c).

%%%%%%%%%%%%%%%%%%%%%%%%%%%%%%%%%%%%%%%%%%%%%%%%%%%%%%
\section{DSF of the $S=1/2$ two-leg spin ladders}
\label{sec:DSFofLADDER}
%%%%%%%%%%%%%%%%%%%%%%%%%%%%%%%%%%%%%%%%%%%%%%%%%%%%%%
%
%%%%%%%%%%%%%%%%%%%%%%%%%%%%%%%%%%%%%%%%%%%%%%%%%%%%%%%%%%%
\begin{figure}[h]
\begin{center}
\includegraphics{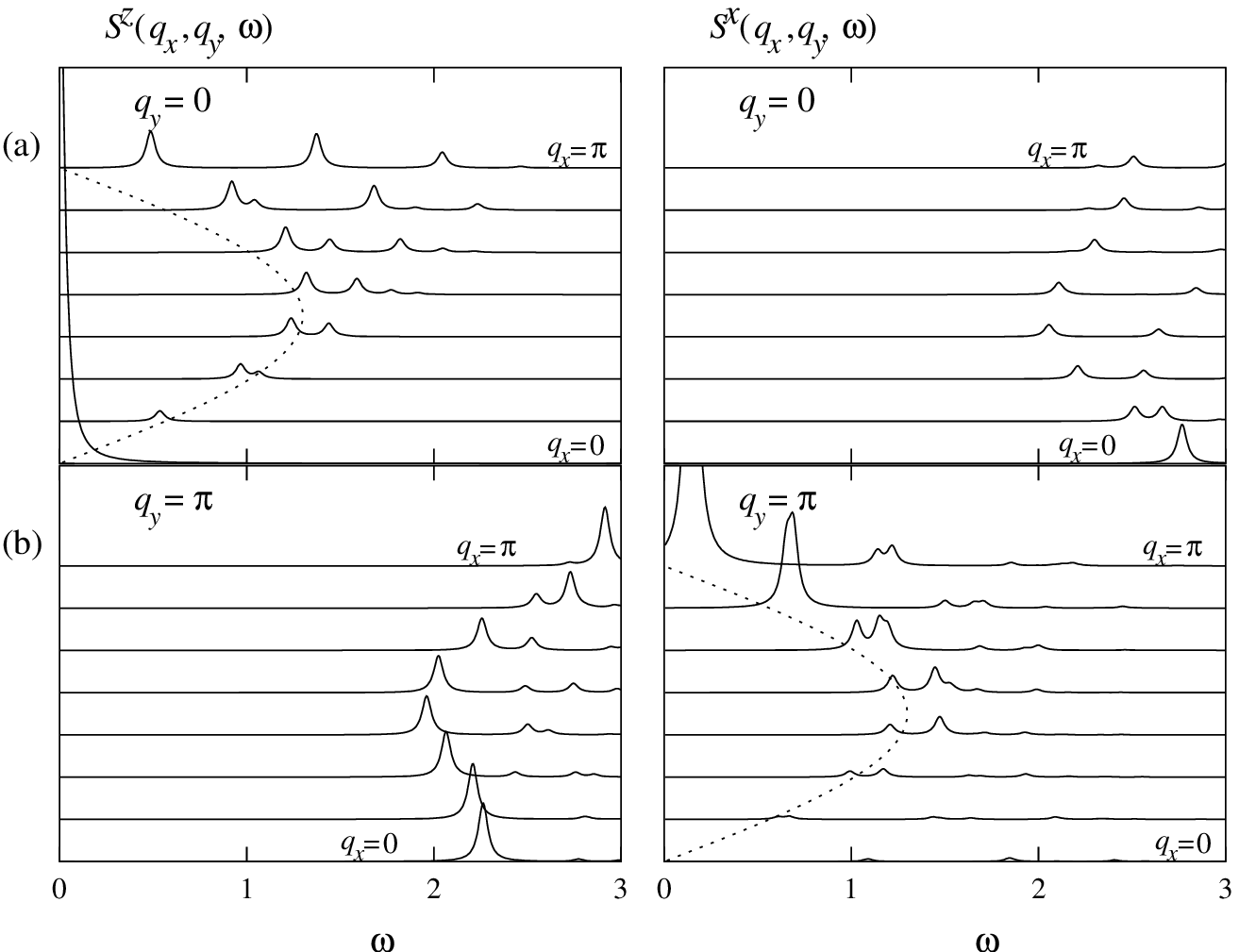}
%\epsfile{file=figure3,height=11cm}
%\figureheight{11cm}
\end{center}
\caption{The DSF of the $S=1/2$ two-leg spin ladder with (a) $q_y=0$ and (b) $q_y=\pi$ is shown for $m=1/4$. The corresponding magnetic field is $H \sim 2.768$. Parameters are $J_1=2.0, J_2=0$, and $J_3=1.0$.}
\label{fig:ladder}
\end{figure}
%%%%%%%%%%%%%%%%%%%%%%%%%%%%%%%%%%%%%%%%%%%%%%%%%%%%%%%%%%%
%
We show the DSF of the $S=1/2$ two-leg spin ladder system for $m=1/4$ with $(N,M)=(28,7)$ in Fig.~\ref{fig:ladder}. 
Parameters are $J_1=2.0, J_2=0$ and $J_3=1.0$.  Thus, the system is in the strong interchain-coupling regime. The wave vector along the rung is fixed to be $q_y=0$ and $\pi$.
The magnetic field corresponding to $m=1/4$ is evaluated as $H \sim 2.768$ from eqs. (\ref{eq:jiba1}) and (\ref{eq:jiba2}). In Fig.~\ref{fig:ladder}, $S^z(q_x,0,\omega)$ and $S^x(q_x,\pi,\omega)$ are gapless, while $S^z(q_x,\pi,\omega)$ and $S^x(q_x,0,\omega)$ are gapful. 
This gapless and gapful behavior is consistent with the asymptotics of the spin correlation functions obtained by the hard-core boson representation of the strong-coupling two-leg ladder~\cite{FZ99}: algebraically decaying ones, 
$\langle S^z_{i,0}S^z_{j,0} \rangle$ and 
$\langle S^x_{i,\pi}S^x_{j,\pi} \rangle$, 
and exponentially decaying ones, 
$\langle S^z_{i,\pi}S^z_{j,\pi} \rangle$ and 
$\langle S^x_{i,0}S^x_{j,0} \rangle$.~\cite{com3}

The strong-coupling $S=1/2$ two-leg spin ladder in magnetic field can be mapped onto the $S=1/2$ $XXZ$ chain with the anisotropy $\Delta=1/2$ in effective magnetic field~\cite{CJFHLBHP,Mila98,GT99,FZ99}. 
As shown in Fig.~\ref{fig:ladder} (a), the intensity of $S^z(q_x,0,\omega)$ lies almost between the boundaries $\omega = (3 \sqrt{3} J_3/4)|\sin q|$ and $\omega = (3 \sqrt{3} J_3/2)|\sin q/2|$, which are the lower and the upper bounds of the excitation continuum of the $S=1/2$ $XXZ$ chain with $\Delta=1/2$.~\cite{MT} 
In $S^x(q_x,\pi,\omega)$, the intensity becomes larger along the lower boundary as  $q_x \rightarrow \pi$. 
The magnetic field corresponding to $m=1/4$ is given as $H_{1/4} = J_1 + J_3/2 = 2.5$ from the effective $S=1/2$ $XXZ$ chain with $\Delta=1/2$~\cite{com1}. 
The value $H_{1/4}=2.5$ is close to our numerical estimation $H \sim 2.768$. 

In $S^z(q_x,\pi,\omega)$ and $S^x(q_x,0,\omega)$, the lower boundaries in the $\omega-q_x$ plane show convexly downward and take the minima at $q_x=\pi/2$. 
The largest intensity of $S^z(q_x,\pi,\omega)$ appears at $q_x=\pi$, while that of $S^x(q_x,0,\omega)$ appears at $q_x=0$. 
The origin of the largest intensity can be explained by the fact that the two modes intersect at $q_x=\pi$ in $S^z(q_x,\pi,\omega)$, and those intersect at $q_x=0$ in $S^x(q_x,0,\omega)$ as seen in Fig.~\ref{fig:ladder} . 
These results are qualitatively the same as those obtained in ref. 14, where the DSF of the weak-coupling $S=1/2$ two-leg spin ladder in $0<m<1/2$~\cite{com2} was calculated analytically using bosonization technique. 
According to ref. 14, furthermore, it is argued that these largest intensities in the gapful DSF come out at $(q_x, \omega) = (\pi, H)$ in $S^z(q_x,\pi,\omega)$ and at $(q_x, \omega) = (0, H)$ in $S^x(q_x,0,\omega)$. From our numerical results, we evaluate that $\omega \sim H( \sim 2.768)$ at both points in agreement with the results in ref. 14. 
It is interesting that the bosonization analysis for the weak-coupling regime well describes the feature of the strong-coupling regime.

\begin{figure}[h]
\begin{center}
\includegraphics{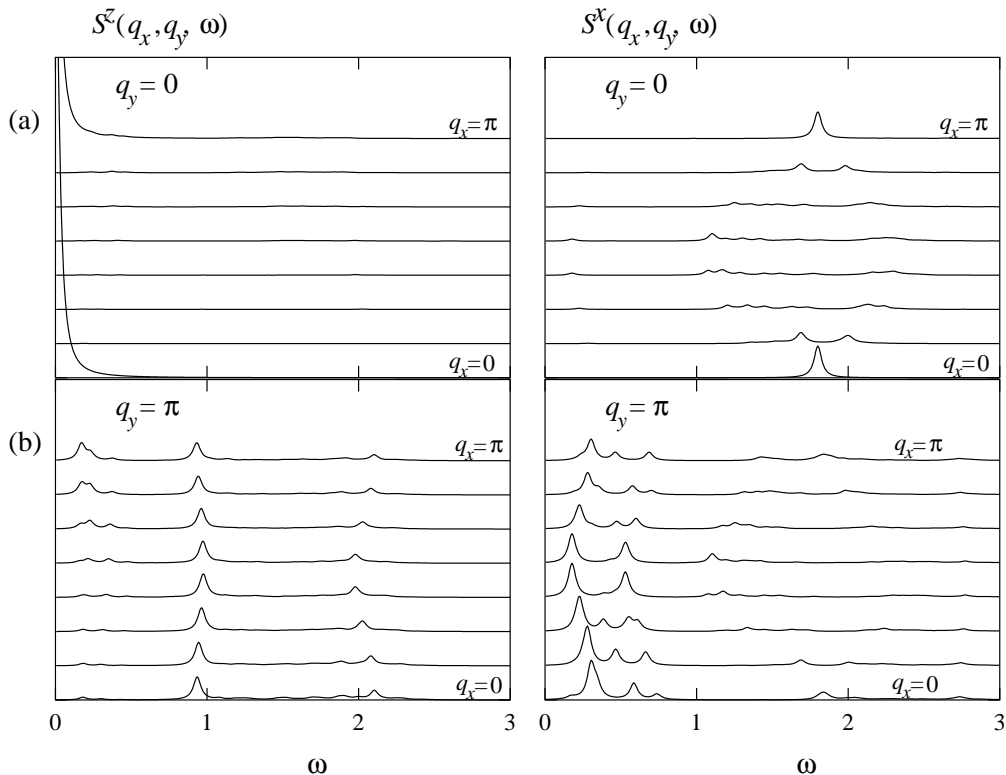}
%\epsfile{file=figure4,height=11cm}
%\figureheight{11cm}
\end{center}
\caption{The DSF of the $S=1/2$ two-leg spin ladder with a diagonal
interaction with (a) $q_y=0$ and (b) $q_y=\pi$ is shown for $m=1/4$.
Parameters are $J_1=1.2, J_2=0.8$, and $J_3=0.4$,  
and the magnetic field is $H=1.80$.}
\label{fig:lwd}
\end{figure}

Finally, we show the DSF of the $S=1/2$ two-leg ladder with a diagonal interaction in comparison with that of the $S=1/2$ bond-alternating spin chain with a NNN interaction. 
Used parameters are $J_1=1.2, J_2=0.8$, and $J_3=0.4$, which are the same as those in Fig.~\ref{fig:chain}(a). We set $m=1/4$ with $(N,M)=(28,7)$. 
The wave vector $q_x=\pi$ in the two-leg ladder geometry corresponds to the wave vector $q=\pi/2$ in the one-dimensional geometry. 
In Fig.~\ref{fig:lwd}, two divergent peaks appear at $q_x=0$ and $\pi$ in $\omega \rightarrow 0$ in $S^z(q_x,0,\omega)$, indicating broken translational symmetry. 
In $S^z (q_x,\pi,\omega)$, the small dispersionless peaks appear around $\omega \sim 1.0$ and $2.0$ as also seen in Fig.~\ref{fig:chain}(a), while in $S^x (q_x,\pi,\omega)$, the peaks appear in $\omega < 1$. 
Since there are no peaks around $\omega \sim 1.0$ and $2.0$ in $S^z (q_x,0,\omega)$, and in $\omega < 1$ in $S^x (q_x,0,\omega)$, these peaks in Fig.~\ref{fig:lwd}(b) are governed by the interchain mode $q_y=\pi$. This is confirmed by the observation that the intensity corresponding to these peaks becomes so weak as $q$ approaches $0$ in Fig.~\ref{fig:chain}(a). 
%
%%%%%%%%%%%%%%%%%%%%%%%%%%%%%%%%%%%%%%%%%%%%%%%%%%%%%%%%%%%
%%%%%%%%%%%%%%%%%%%%%%%%%%%%%%%%%%%%%%%%%%%%%%%%%%%%%%%%%%%
%

%%%%%%%%%%%%%%%%%%%%%%%%%%%%%%%%%%%%%%%%%%%%%%%%%%%%%%
\section{Summary}
\label{sec:Summary}
%%%%%%%%%%%%%%%%%%%%%%%%%%%%%%%%%%%%%%%%%%%%%%%%%%%%%%
We have calculated the DSF of the $S=1/2$ bond-alternating spin chain with a NNN interaction in magnetic fields, using the continued fraction method based on the Lanczos algorithm. 
When the plateau exists on the magnetization curve, large intensity with a periodic dispersion relation comes out in $S^z(q,\omega)$, whose periodicity and amplitude are sensitive to the coupling constants. On the other hand, large intensity with an almost dispersionless mode comes out in $S^x(q,\omega)$. 
When the plateau does not exist, the DSF shows qualitatively the same behavior as that of the $S=1/2$ Heisenberg chain. 
 
The DSF of the $S=1/2$ two-leg ladder for $m=1/4$ is also calculated in the strong interchain-coupling regime. 
 The DSF shows gapless or gapful behavior depending on the wave vector $q_y$. The gapless modes in $S^z(q_x,0,\omega)$ and $S^x(q_x,\pi,\omega)$ are well described by the $S=1/2$ $XXZ$ chain with $\Delta=1/2$. The gapful behavior in $S^z(q_x,\pi,\omega)$ and $S^x(q_x,0,\omega)$ is consistent with the results based on bosonization technique for the weak-coupling regime.

%%%%%%%%%%%%%%%%%%%%%%%%%%%%%%%%%%%%%%%%%%%%%%%%%%%%%%
\section*{Acknowledgments}
%%%%%%%%%%%%%%%%%%%%%%%%%%%%%%%%%%%%%%%%%%%%%%%%%%%%%%
The authors would like to thank Professor T. Tonegawa for sending numerical data. One of authors (S. S.) would like to thank Professor I. Harada for sending preprints. Numerical computations were carried out at the computer facility of Yukawa Institute for Theoretical Physics. This work was supported by the Grant-in-Aid no. 10640344 for Scientific Research from the Ministry of Education, Science, Sports and Culture, Japan.

%%%%%%%%%%%%%%%%%%%%%%%%%%%%%%%%%%%%%%%%%%%%%%%%%%%%%%%%%%%%%%%%%%%%%
%                        REFERENCES                                 %
%%%%%%%%%%%%%%%%%%%%%%%%%%%%%%%%%%%%%%%%%%%%%%%%%%%%%%%%%%%%%%%%%%%%%
%

%
%
%%%%%%%%%%%%%%%%%%%%%%%%%%%%%%%%%%%%%%%%%%%%%%%%%%%%%%%%%%%%%%%%%%%%%


\begin{thebibliography}{99}
\bibitem{com3} 
In our expression, $S_{i,q_y}^\mu = (1/\sqrt{2}) \sum_{l=1,2} {\rm e}^{i q_y l} S_{l,j}^\mu$ $(\mu=x,y,z)$. 
Thus, $S_{i,0}^\mu = (1/\sqrt{2}) (S_{1,j}^\mu + S_{2,j}^\mu)$ and 
$S_{i,\pi}^\mu = (1/\sqrt{2}) (-S_{1,j}^\mu + S_{2,j}^\mu)$. 
The suffix $(i,q_y)$ for $S^{\mu}_{i,q_y}$ is interchanged as compared with that in ref. 14.
\bibitem{com1} 
In the effective $S=1/2$ $XXZ$ chain with $\Delta=1/2$, the effective magnetic field $\tilde{H}$ and magnetization $\tilde{m}$ are described as $\tilde{H} = H - J_1 - J_3/2$ and $\tilde{m} = 2m - 1/2$ by using parameters and $m$ of the original Hamiltonian. Therefore, when $m=1/4$, $\tilde{m} = 0$ and then $\tilde{H} = 0$, leading to $H = J_1 + J_3/2$.
\bibitem{com2}
In ref. 14, the saturated magnetization is set to be $1$. Thus we have to use $m=1/2$ in the results of ref. 14 to compare the present results.
\end{thebibliography}
\end{document}